\documentclass[conference]{IEEEtran}
\IEEEoverridecommandlockouts

\usepackage{cite}
\usepackage{amsmath,amssymb,amsfonts}
\usepackage{algorithmic}
\usepackage{graphicx}
\usepackage{textcomp}
\usepackage{xcolor}
\usepackage{booktabs}
\usepackage{multirow}
\usepackage{colortbl}
\usepackage{subcaption}
\usepackage{hyperref}

\def\BibTeX{{\rm B\kern-.05em{\sc i\kern-.025em b}\kern-.08em
    T\kern-.1667em\lower.7ex\hbox{E}\kern-.125emX}}

\begin{document}

\title{\fontsize{23}{25}\selectfont Compress the Cache, Not the Speech Embedding: \\KV Compression for Efficient Speech LLMs}

\author{
  \IEEEauthorblockN{Ke-Han Lu$^{12*}$\thanks{$^*$Work done while Ke-Han Lu was an intern at Microsoft.}, Keqi Deng$^{1\dagger}$\thanks{$^\dagger$Corresponding author.}, Ruchao Fan$^1$, Rui Zhao$^1$, Jinyu Li$^1$}
  \IEEEauthorblockA{
    {$^1$Microsoft}, USA \\
    {$^2$National Taiwan University}, Taiwan\\
    d12942024@ntu.edu.tw, keqideng@microsoft.com
  }
}


\maketitle

\begin{abstract}

Speech large language models (Speech LLMs) typically encode speech into sequences far longer than text, creating a major efficiency bottleneck during autoregressive decoding.
A common remedy is to compress the speech sequence at the adapter level to remove temporal redundancy before it enters the LLM; however, such early downsampling risks discarding fine-grained information that cannot be recovered.
We propose \textit{SpeechKV}, which applies a learned pooling to the KV cache of speech tokens inside the LLM. This design allows the LLM to fuse speech and text internally while directly accelerating decoding.
Trained on 71K hours of speech data, \textit{SpeechKV} compresses the speech to approximately text-level granularity yet maintains performance on par with or even slightly better than the uncompressed baseline, with relative gains of 6.6\% on out-of-domain entity recognition and 2.3\% on OpenASR, while delivering at least 1.49 times decoding speedup that scales with audio length.

\end{abstract}

\begin{IEEEkeywords}
Speech LLM, KV cache compression, automatic speech recognition
\end{IEEEkeywords}

\section{Introduction}
\begin{figure*}[t]
    \centering
    \includegraphics[width=\linewidth]{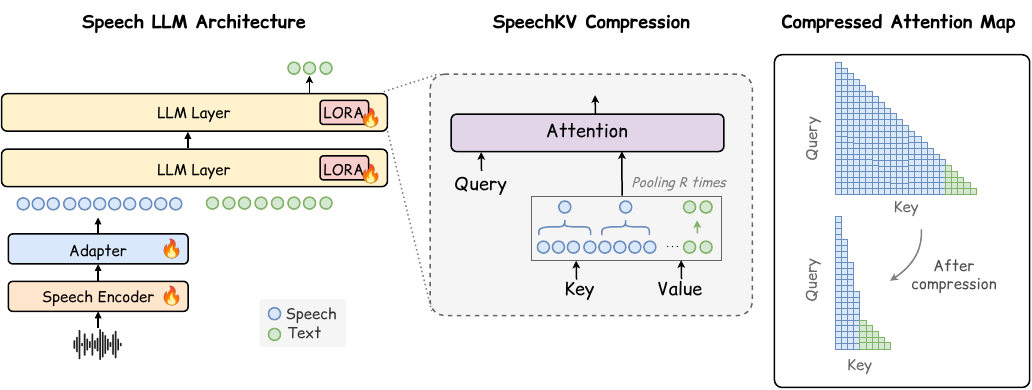}
    \caption{Overview of our approach. \textbf{Left:} the Speech LLM follows the encoder--adapter--LLM paradigm. We apply the KV cache compression in deeper layers of the LLM. \textbf{Middle:} from layer $l_0$ onward, the learned pooling compresses each window of speech keys and values into a single representation, while text positions remain unchanged. \textbf{Right:} In both training and decoding stage, the full-resolution queries attend to a shorter sequence of keys.}
    \label{fig:overview}
\end{figure*}

Large language models (LLMs) have demonstrated remarkable capabilities across a wide range of natural language processing tasks~\cite{peng2023instruction, dubey2024llama,yang2025qwen3,abdin2024phi}. Building on this success, researchers have extended these models to speech modalities, giving rise to a growing family of Speech LLMs~\cite{tang2024salmonn,yang2024qwen2technicalreport,11077996,arora2025landscapespokenlanguagemodels,abouelenin2025phi,lu24c_interspeech,desta2,lu2025desta25,10447605,hu2024wavllm,ma2024embarrassingly,deng-etal-2025-wav2prompt,10445874,sun2026speech,deng2026speech,10389705}. A typical Speech LLM consists of three components: a pre-trained speech encoder, a  modality adapter, and a large language model backbone. The speech encoder maps input speech into a sequence of acoustic frames. The modality adapter projects these frames into the LLM's input space, and the LLM then processes speech embeddings alongside text embeddings to enable multimodal understanding.

A key challenge in Speech LLMs is the length of speech sequences. For example, in English ASR, a single text token typically corresponds to approximately 300\,ms of speech, yet common speech encoders emit one frame every 80\,ms, yielding roughly 4$\times$ more sequence length for the same semantic information. This poses a substantial efficiency bottleneck during autoregressive generation, where the model must attend to all preceding positions at each decoding step, resulting in increased latency and memory consumption that scale with sequence length.

A straightforward approach is to downsample speech sequences at the adapter level into more compact representations\cite{11077996, 10389705, 10445874, deng-etal-2025-wav2prompt,ma2024embarrassingly,deng2026speech,mohapatra2026speechmapper}. However, such early compression may risk losing information and fails to exploit the representational capacity of the LLM itself. This creates a trade-off between efficiency and recognition performance.

In this work, we aim at resolving this trade-off by deferring speech compression into the LLM itself. Recent LLM research has shown that Transformer layers often produce redundant key-value representations~\cite{NEURIPS2024_28ab4182,3737916.3742359}. In our preliminary experiments, we also observe that Speech LLMs naturally merge nearby speech information in their early layers and operate on near-duplicate representations in the deeper layers (see Section~\ref{subsec:redundancy}).
Based on this observation, we propose \textit{\textbf{SpeechKV}}, which applies a learned pooling operation to merge the speech key-value sequences at an intermediate LLM layer as illustrated in Figure~\ref{fig:overview}. This allows the LLM layer to learn which speech features to preserve and how to fuse speech with text through the attention mechanism.

We evaluate SpeechKV on a Speech LLM with Qwen3-1.7B backbone trained on 71K hours of ASR data. Interestingly, when compressing the KV sequences close to text granularity, SpeechKV stays on par with the uncompressed baseline and even slightly surpasses it, with relative gains of 6.6\% on out-of-domain entity recognition and 2.3\% on the OpenASR benchmark~\cite{srivastav2025openasrleaderboardreproducible}, while achieving at least $1.49\times$ decoding speedup with vLLM~\cite{kwon2023efficient}.
Further analysis of the attention patterns shows that, with SpeechKV, the LLM's deeper layers allocate more focused attention rather than spreading attention across redundant keys. This suggests that SpeechKV not only accelerates decoding but also provides a beneficial regularization for the speech-text alignment.

Our contributions are summarized as follows:
\begin{itemize}
  \item We propose SpeechKV, a method that applies learned pooling to compress the speech KV cache at an intermediate Speech LLM layer.
  \item We demonstrate compressing speech to text-level granularity maintains the performance of the uncompressed baseline while achieving $1.49$ to $2\times$ decoding speedup.
  \item We empirically show that adjacent speech representations become increasingly redundant across Speech LLM layers, providing the motivation for deferring compression into the LLM rather than applying it at the adapter level.
\end{itemize}

\section{Related Work}


\subsection{Speech LLMs and Speech Sequence Compression}

Motivated by the advancements in text-centric large language models (LLMs), researchers have actively explored extending these architectures to process audio and spoken language, giving rise to a rapidly growing family of Speech LLMs. These models typically adopt an encoder--adapter--LLM paradigm, which enables the LLM to directly handle speech inputs.
Within this framework, the central challenge is effectively bridging the modality and length discrepancy between continuous speech and discrete text.

The full encoder output can be simply retained via
a linear projector or convolution layers~\cite{ma2024embarrassingly,yang2024qwen2technicalreport,abouelenin2025phi}, but the resulting long speech sequences drastically increase computational cost during LLM decoding. To address this, most existing works perform pre-LLM compression through specialized adapters that reduce the sequence length before it enters the LLM. Representative approaches include CTC-based boundary merging or blank filtering~\cite{10.1145/1143844.1143891,11077996, 10888940, deng2026speech,verdini2024connect}, CIF-based \cite{dong2020cif} dynamic accumulation of acoustic features~\cite{deng-etal-2025-wav2prompt,verdini2024connect}, and Q-Formers that extract a fixed number of speech representations~\cite{pmlr-v202-li23q,11077996,10445874,verdini2024connect,held2024distilling,tang2024salmonn}. The LLM then processes these down-sampled speech prompts along with the text sequence. While effective, these approaches introduce a bottleneck that risks an unrecoverable loss of information in the subsequent LLM layers.
This motivates an alternative direction that compresses the speech KV cache inside the LLM itself, where attention layers can still access full-resolution queries and residual streams.

\subsection{KV Cache Compression in LLMs}

The linearly growing KV cache is a well-known memory and latency bottleneck for long-context LLM inference~\cite{li2025a}. Several studies have explored training-free approaches that manage the KV cache during the inference stage. For example, eviction-based methods retain only a subset of cached tokens based on attention statistics or recency \cite{NEURIPS2023_6ceefa7b, ICLR2024_5e5fd18f, NEURIPS2024_28ab4182}, while other approaches utilize low-bit quantization of cached states \cite{NEURIPS2024_028fcbcf} to reduce per-token memory cost. However, these training-free methods inevitably suffer from a train-test mismatch, since the model has never been exposed to truncated or evicted cache patterns during training.
Trainable approaches instead learn the compression jointly with the model, for example by merging or summarizing context into a shorter KV cache on the fly \cite{nawrot2024dynamic, wang2024model, chevalier-etal-2023-adapting, sun2026speech}. Yet it remains underexplored how a Speech LLM internally processes speech sequences across layers and whether this behavior can inform a simpler compression strategy. In this work, we analyze layer-wise attention behavior and use the findings to guide speech KV cache compression.

\section{Method}
\label{sec:method}

\subsection{Speech Language Model}
\label{sec:speech_llm}

Our baseline model follows the encoder--adapter--LLM paradigm illustrated in Figure~\ref{fig:overview}.
A speech encoder $f_{\mathrm{enc}}$ converts the input speech $\mathbf{x}$ into frame-level acoustic features:
\begin{equation}
  \mathbf{Z} = f_{\mathrm{enc}}(\mathbf{x}) \in \mathbb{R}^{T \times d_e},
\end{equation}
where $T$ is the number of frames and $d_e$ is the encoder hidden dimension.
A 2-layer MLP adapter projects these frames into the LLM embedding space without temporal downsampling:
\begin{equation}
  \mathbf{A} = f_{\mathrm{adp}}(\mathbf{Z}) \in \mathbb{R}^{T \times d}.
\end{equation}
Together with the text embeddings $\mathbf{E} \in \mathbb{R}^{N \times d}$, the concatenated sequence $[\mathbf{A};\;\mathbf{E}]$ is fed into a decoder-only LLM with $L$ layers as the input and trained with next-token prediction to generate $M$-token outputs:
\begin{equation}
  \mathcal{L} = -\sum_{t=1}^{M} \log\, p(y_t \mid \mathbf{A}, \mathbf{E}, y_{<t}).
\end{equation}
Since the speech sequence is typically much longer than the text portion ($T \gg N + M$), the speech KV cache dominates per-step attention cost during autoregressive decoding, forming the primary inference bottleneck.

\subsection{SpeechKV}
\label{sec:compression}

As illustrated in Figure~\ref{fig:overview}, we propose to compress the speech KV sequences inside the LLM to accelerate decoding. Starting from layer $l_0$, we apply a learned pooling operator $\mathcal{P}_R$ that compresses the speech keys and values into a shorter sequence with compression ratio $R$.

Specifically, the operator $\mathcal{P}_R$ applies only on the $T$ speech positions and leaves the text positions unchanged. It partitions the speech sequences into non-overlapping uniform windows of size $R$. Within each window, a learned linear gate $g: \mathbb{R}^{d_h} \to \mathbb{R}$ produces a scalar logit from each key vector $\mathbf{k}_j \in \mathbb{R}^{d_h}$, and a softmax normalizes these logits into per-position weights:
\begin{equation}
  \alpha_j = \frac{\exp\bigl(g(\mathbf{k}_j)\bigr)}{\sum_{j'=1}^{R} \exp\bigl(g(\mathbf{k}_{j'})\bigr)}, \quad
  \hat{\mathbf{k}} = \sum_{j=1}^{R} \alpha_j\, \mathbf{k}_j, \quad
  \hat{\mathbf{v}} = \sum_{j=1}^{R} \alpha_j\, \mathbf{v}_j.
\end{equation}
The gate is initialized to zero so that the model starts from uniform averaging and gradually learns to emphasize informative positions within each window. 

At each LLM layer $l \geq l_0$, the full-resolution queries attend over the compressed KV sequence:
\begin{equation}
  \mathrm{Attention}^{(l)} = \mathrm{softmax}\!\left(\frac{\mathbf{Q}^{(l)} \hat{\mathbf{K}}^{(l)\top}}{\sqrt{d_h}}\right) \hat{\mathbf{V}}^{(l)},
\end{equation}
where $\hat{\mathbf{K}}^{(l)} = \mathcal{P}_R(\mathbf{K}^{(l)})$ and $\hat{\mathbf{V}}^{(l)} = \mathcal{P}_R(\mathbf{V}^{(l)})$. The attention mask is correspondingly adjusted so that each query attends only to its preceding positions, preserving causality over the compressed sequence, as illustrated in Figure~\ref{fig:overview}(Right). 

We intentionally keep the compression operation simple and lightweight, relying on the LLM's own attention mechanism to effectively process the compressed sequences with negligible overhead.

\subsection{Training and Inference}
\label{sec:train_infer}

During training, SpeechKV is applied on-the-fly within the forward pass of each target layer and jointly optimized with next-token-prediction loss, ensuring no train-test mismatch. 
In the inference stage, the compression is performed once during prefill, and the compressed KV is stored in the cache, yielding an $R$-fold reduction in speech KV cache size at each decoding step. This directly translates to a smaller memory footprint and faster decoding speed.


\begin{table}[t]
\centering
\caption{Training data composition (71K hours total).}
\label{tab:training_data}
\begin{tabular}{lrr}
\toprule
\textbf{Dataset} & \textbf{Hours} & \textbf{Utts (K)} \\
\midrule
Multilingual LibriSpeech (en)~\cite{Pratap2020MLSAL}  & 44,659.7 & 10,808.0 \\
GigaSpeech~\cite{GigaSpeech2021}           & 13,714.4 & 11,659.8 \\
SPGISpeech~\cite{2021arXiv210402014O}         &  4,996.6 &  1,965.3 \\
LibriSpeech~\cite{panayotov2015librispeech}    &  1,920.0 &    560.8 \\
LibriHeavy~\cite{kang2023libriheavy}          &  1,852.2 &    469.4 \\
Common Voice 15~\cite{commonvoice:2020}   &  1,688.3 &  1,069.7 \\
VoxPopuli~\cite{wang-etal-2021-voxpopuli}             &  1,045.2 &    365.0 \\
TED-LIUM 3~\cite{hernandez2018ted}       &    907.6 &    536.5 \\
Earnings-22~\cite{delrio2022earnings22}        &    287.2 &    104.0 \\
AMI~\cite{10.1007/11677482_3}                     &    156.3 &    217.0 \\
FLEURS~\cite{fleurs2022arxiv}               &     15.0 &      5.2 \\
\midrule
\textbf{Total}                                 & \textbf{71,242.5} & \textbf{27,760.8} \\
\bottomrule
\end{tabular}
\end{table}


\begin{table*}[t]
\centering
\caption{Comparison of compression methods at $R{=}4\times$ and $R{=}8\times$. Our in-LLM methods (HS Compression, SpeechKV) apply $l_0=5$. We report entity error rate (EER) for entity recognition and word error rate (WER) for ASR task, respectively.}
\label{tab:main}
\setlength{\tabcolsep}{3pt}
\begin{tabular}{l ccc ccccc cccccccc}
\toprule
\multirow{3}{*}{\textbf{Method}} & \multicolumn{3}{c}{\textbf{In-houseER} $\downarrow$} & \multicolumn{5}{c}{\textbf{In-houseASR} $\downarrow$} & \multicolumn{8}{c}{\textbf{OpenASR} $\downarrow$} \\
\cmidrule(lr){2-4} \cmidrule(lr){5-9} \cmidrule(lr){10-17}
 & \cellcolor{gray!40}AVG & Banking & Medical & \cellcolor{gray!40}AVG & Dict. & Fin.Call & Brod. & VoiceM. & \cellcolor{gray!40}AVG & AMI & Earn22 & Giga & LS-c & LS-o & SPGi & Vox \\
\midrule
\textit{Open-sourced Models} \\
Whisper-L-v3\cite{radford2023robust} & 8.41 & 9.43 & 7.38 & 4.91 & 3.40 & 6.83 & 4.58 & 4.82 & 7.95 & 15.95 & 11.29 & 10.02 & 2.01 & 3.91 & 2.94 & 9.54 \\
Phi4-mm\cite{abouelenin2025phi} & 11.36 & 13.27 & 9.45 & 4.89 & 3.13 & 6.21 & 3.88 & 6.33 & 6.60 & 11.69 & 10.16 & 9.78 & 1.68 & 3.83 & 3.13 & 5.91 \\

\midrule
MLP (Baseline) & 14.41 & 16.67 & 12.14 & 5.78 & 3.59 & 7.58 & 4.26 & 7.67 & 6.12 & 10.59 & 8.59 & 9.44 & 1.83 & 4.03 & 2.02 & 6.36 \\
\midrule
\textit{$R=4$} \\
Concat-MLP      & 15.84 & 17.76 & 13.91 & 6.29 & 4.66 & 7.45 & 5.14 & 7.91 & 6.07 & 9.65 & 8.71 & 9.55 & 1.86 & 4.09 & 2.08 & 6.52 \\
Window Q-Former & 14.70 & 17.32 & 12.07 & 6.23 & 4.91 & 7.39 & 4.98 & 7.65 & 6.02 & \textbf{9.28} & 8.93 & 9.54 & 1.85 & 4.07 & 2.04 & 6.43 \\
HS Compression      & 15.22 & 17.54 & 12.91 & 5.79 & \textbf{3.62} & 7.15 & 4.67 & 7.73 & 6.06 & 9.53 & 8.85 & 9.54 & 1.85 & 4.10 & 2.08 & 6.44 \\
SpeechKV       & \textbf{13.46} & \textbf{15.30} & \textbf{11.62} & \textbf{5.71} & 3.83 & \textbf{7.09} & \textbf{4.52} & \textbf{7.42} & \textbf{5.98} & 9.68 & \textbf{8.60} & \textbf{9.37} & \textbf{1.79} & \textbf{4.05} & \textbf{2.03} & \textbf{6.36} \\
\midrule
\textit{$R=8$} \\
Concat-MLP      & 17.52 & 20.94 & 14.11 & 6.63 & 4.08 & 7.63 & 6.17 & 8.63 & 6.43 & 10.24 & 9.27 & 9.95 & 2.00 & 4.43 & 2.22 & 6.92 \\
Window Q-Former & 16.41 & 18.80 & 14.01 & 6.50 & 4.34 & 7.59 & 5.08 & 8.98 & 6.64 & 12.02 & 9.20 & 9.96 & 1.91 & 4.48 & 2.20 & 6.72 \\
HS Compression      & 16.98 & 19.46 & 14.49 & 6.22 & 4.06 & 7.57 & 4.77 & 8.47 & 6.31 & 9.80 & 9.34 & 9.85 & 1.94 & 4.33 & 2.23 & 6.66 \\
SpeechKV       & \textbf{14.91} & \textbf{17.00} & \textbf{12.82} & \textbf{5.95} & \textbf{3.95} & \textbf{7.27} & \textbf{4.45} & \textbf{8.14} & \textbf{6.05} & \textbf{9.37} & \textbf{8.84} & \textbf{9.59} & \textbf{1.86} & \textbf{4.12} & \textbf{2.13} & \textbf{6.45} \\
\bottomrule
\end{tabular}%
\end{table*}

\section{Experiment Setup}

\subsection{Model}
The backbone language model is Qwen3-1.7B~\cite{yang2025qwen3}, comprising 28 Transformer decoder layers. The speech encoder is initialized from a pretrained encoder-decoder ASR model which consists of 3 convolution layers followed by 24 conformer blocks~\cite{gulati20_interspeech}, producing speech frames at an 80\,ms frame rate. In the uncompressed \textbf{MLP Baseline}, a 2-layer MLP projects the encoder output directly into the LLM input embedding space without temporal downsampling.

For compressed settings, we uniformly merge every $R$ consecutive frames into a single speech position, so that each resulting position corresponds to $80R$\,ms of speech. We compare compression strategies at two levels.
At the adapter level, we study two variants:
(1)~\textbf{Concat-MLP} concatenates $R$ adjacent frames along the feature dimension and projects the result through a 2-layer MLP;
(2)~\textbf{Window Q-Former}~\cite{pmlr-v202-li23q} employs a 2-layer Transformer cross-attention module to attend over $R$ adjacent frames and produce one merged speech embedding.

At the LLM level, we keep the uncompressed 2-layer MLP as the adapter and instead perform compression inside the LLM. Based on the empirical observation from Section~\ref{subsec:redundancy}, we use $l_0 = 5$ as default setting. This allows the first 4 layers to process the full-resolution speech sequence.
(3)~\textbf{SpeechKV} compresses only the key and value projections from layer $l_0$ onward, uniformly merging every $R$ consecutive speech KV vectors into one while preserving full-resolution queries and the residual stream.
We additionally propose (4)~\textbf{HS Compression} (hidden-state compression), an alternative in-LLM design that applies the same learned pooling to the entire hidden state at layer $l_0$, thereby shortening all subsequent layers including queries. Comparing (3) and (4) directly isolates the benefit of preserving full-resolution queries in SpeechKV.

\subsection{Training}
We use a combined 71K hours of open-sourced ASR data, as detailed in Table~\ref{tab:training_data}.
During training, the speech encoder, the adapter module, and LoRA~\cite{hulora} with rank 32 are trainable, while the remaining LLM weights are frozen, yielding approximately 480M trainable parameters out of 2.2B in the entire model.
The model is optimized with the AdamW optimizer using a peak learning rate of $2 \times 10^{-5}$, linear warmup over the first 100 steps, and a warmup-decay schedule. Training is conducted with DeepSpeed ZeRO Stage 1. The effective batch size is approximately 200K tokens including speech and text tokens, and each model is trained for 10,000 steps, covering the full training set approximately once. Each training run takes about 1 day on 8 H100 GPUs.

\subsection{Evaluation}
We evaluate SpeechKV on three test suites that capture different aspects of recognition quality.
(1)~\textbf{OpenASR}\footnote{Leaderboard version accessed May 20, 2026.}~\cite{srivastav2025openasrleaderboardreproducible} comprises seven publicly available benchmarks (AMI~\cite{10.1007/11677482_3}, Earnings-22~\cite{delrio2022earnings22}, GigaSpeech~\cite{GigaSpeech2021}, LibriSpeech clean/other~\cite{panayotov2015librispeech}, SPGISpeech~\cite{2021arXiv210402014O}, and VoxPopuli~\cite{wang-etal-2021-voxpopuli}), that serve as an in-domain ASR evaluation.
(2)~\textbf{In-house ASR} consists of four proprietary test sets spanning dictation, financial calls, broadcast news, and voicemail. These sets are out-of-domain with respect to the training data to test the model's generalization ability and robustness.
(3)~\textbf{In-house Entity Recognition (ER)} includes two entity-dense test sets from banking and medical domains.
We report entity error rate (EER), defined as $1-\text{Entity Recall}$, which measures the proportion of entities not correctly recognized. An entity is considered correct only if all of its constituent words are transcribed without error.

\section{Results}

\subsection{Main Results}
\label{subsec:main_results}

Table~\ref{tab:main} compares the proposed SpeechKV against adapter-level approaches (Concat-MLP, Window Q-Former) and our alternative in-LLM method (HS Compression) at two compression ratios. Adapter-level methods compress the speech embeddings before they enter the LLM, while in-LLM methods apply compression at intermediate LLM representations. We also include two open-sourced models, Whisper~\cite{radford2023robust} and Phi4-mm~\cite{abouelenin2025phi}, as external references. Since these models were trained on millions of hours of web-scale speech, they are not directly comparable with our 71K-hour setup.
Overall, our proposed SpeechKV consistently performs best among the compression strategies and maintains baseline performance across most evaluation metrics and compression ratios. 

When comparing methods at $R{=}4$, which aligns the granularity of speech with text tokens, all methods achieve slightly better performance on OpenASR. This suggests that the intensive 71K hours of in-domain training data allow the encoder and LLM to jointly learn to compensate the compression. However, on the out-of-domain test set, performance becomes more sensitive to compression. Concat-MLP and Window Q-Former show degradation on both ER and ASR, while
HS Compression maintains competitive ASR performance but suffers clear ER degradation. This reflects the loss of information from compressing entire speech sequences that requires fine-grained discrimination. In contrast, SpeechKV stays on par with the uncompressed baseline across most evaluation suites, and even surpasses it, with relative gains of 6.6\%, 1.2\%, and 2.3\% on In-house ER, In-house ASR, and OpenASR, respectively.

We conjecture that this behavior arises from two complementary effects. First, because each layer retains access to complete audio information, the model can progressively adjust to a shorter KV cache with minimal information loss.
Second, the compressed granularity aligns with the text-level token rate that the LLM is pretrained to process. Together, these effects act as a structural regularization that encourages more focused speech--text alignment, which explains why SpeechKV maintains or slightly improves the baseline rather than merely approximating it.

The $R{=}8$ results further corroborate this hypothesis. At this aggressive ratio, the performance gap between methods widens considerably: all competing methods degrade by at least 8\% relative on In-house metrics, with some exceeding 20\%. In contrast, SpeechKV degrades gracefully, within 3.5\% relative degradation on In-house ER and 2.9\% on In-house ASR, while still matching the baseline on OpenASR. 
The widening gap confirms our design advantage, demonstrating that SpeechKV is a promising direction for building efficient speech LLMs without sacrificing recognition quality.

\subsection{Layer-wise Analysis and Ablation Studies}
\label{subsec:redundancy}

\begin{figure}
    \centering
    \includegraphics[width=0.98\linewidth]{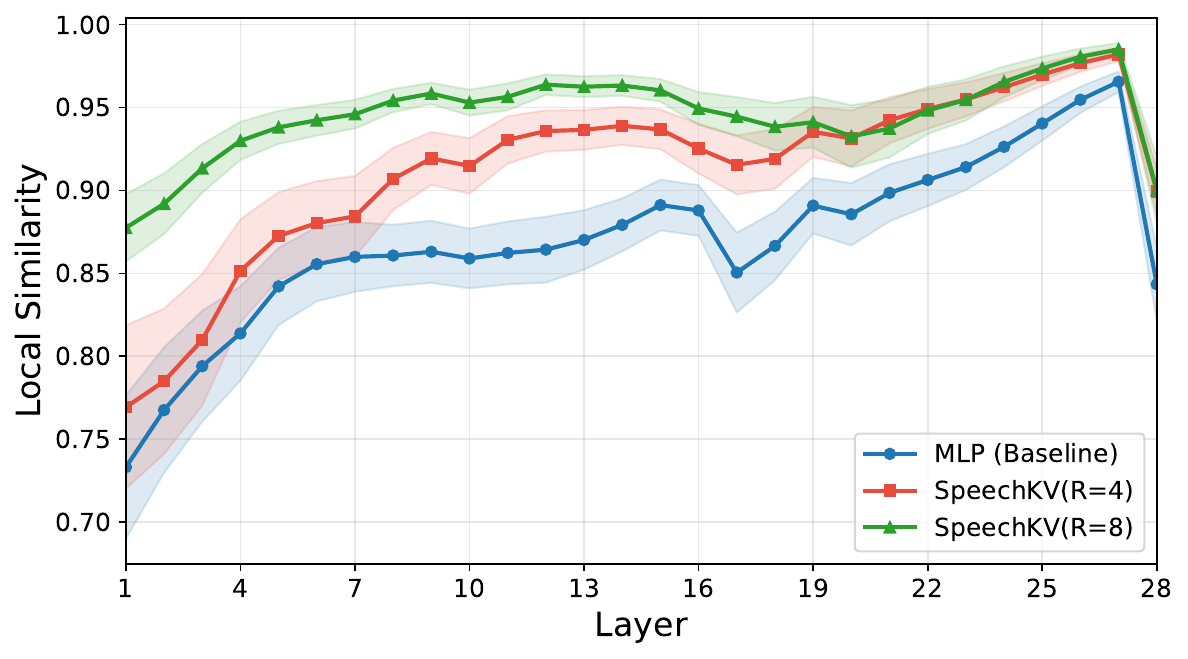}
    \caption{Layer-wise local similarity (cosine similarity between  adjacent speech positions) for the uncompressed baseline and the KV-Compressed models. }
    \label{fig:local_similarity_comparison}
\end{figure}

\begin{table}[t]
\centering
\caption{Ablation results of layer ablation for HS Compression and SpeechKV from different starting layer $l_0$. The results present the average score for each testing set.}
\label{tab:layer_ablation}
\begin{tabular}{ll ccc}
\toprule
\textbf{Method} & \textbf{$l_0$} & \textbf{In-houseER} & \textbf{In-houseASR} & \textbf{OpenASR} \\
\midrule
MLP (Baseline) & -- & 14.41 & 5.78 & 6.12 \\
\midrule
\multicolumn{5}{l}{\textit{$R = 4\times$}} \\

HS Compression & 3 & 15.69 & 6.38 & 6.19 \\
HS Compression & 5 & \textbf{15.22} & \textbf{5.79} & 6.06 \\
HS Compression & 7 & 15.39 & 5.86 & \textbf{6.03} \\
\cmidrule(lr){1-5}
SpeechKV & 3 & 14.33 & 5.87 & 6.12 \\
SpeechKV & 5 & \textbf{13.46} & \textbf{5.71} & 5.98 \\
SpeechKV & 7 & 13.56 & 6.23 & \textbf{5.93} \\
\midrule
\multicolumn{5}{l}{\textit{$R = 8\times$}} \\

HS Compression & 3 & 17.78 & 6.80 & 6.63 \\
HS Compression & 5 & \textbf{16.98} & \textbf{6.22} & 6.31 \\
HS Compression & 7 & 18.12 & 6.60 & \textbf{6.30} \\
\cmidrule(lr){1-5}
SpeechKV & 3 & 16.70 & 7.06 & 6.53 \\
SpeechKV & 5 & \textbf{14.91} & \textbf{5.95} & \textbf{6.05} \\
SpeechKV & 7 & 15.25 & 6.49 & 6.08 \\
\bottomrule
\end{tabular}
\end{table}

\begin{figure}
    \centering
    \begin{subfigure}{0.98\linewidth}
        \centering
        \includegraphics[width=\linewidth]{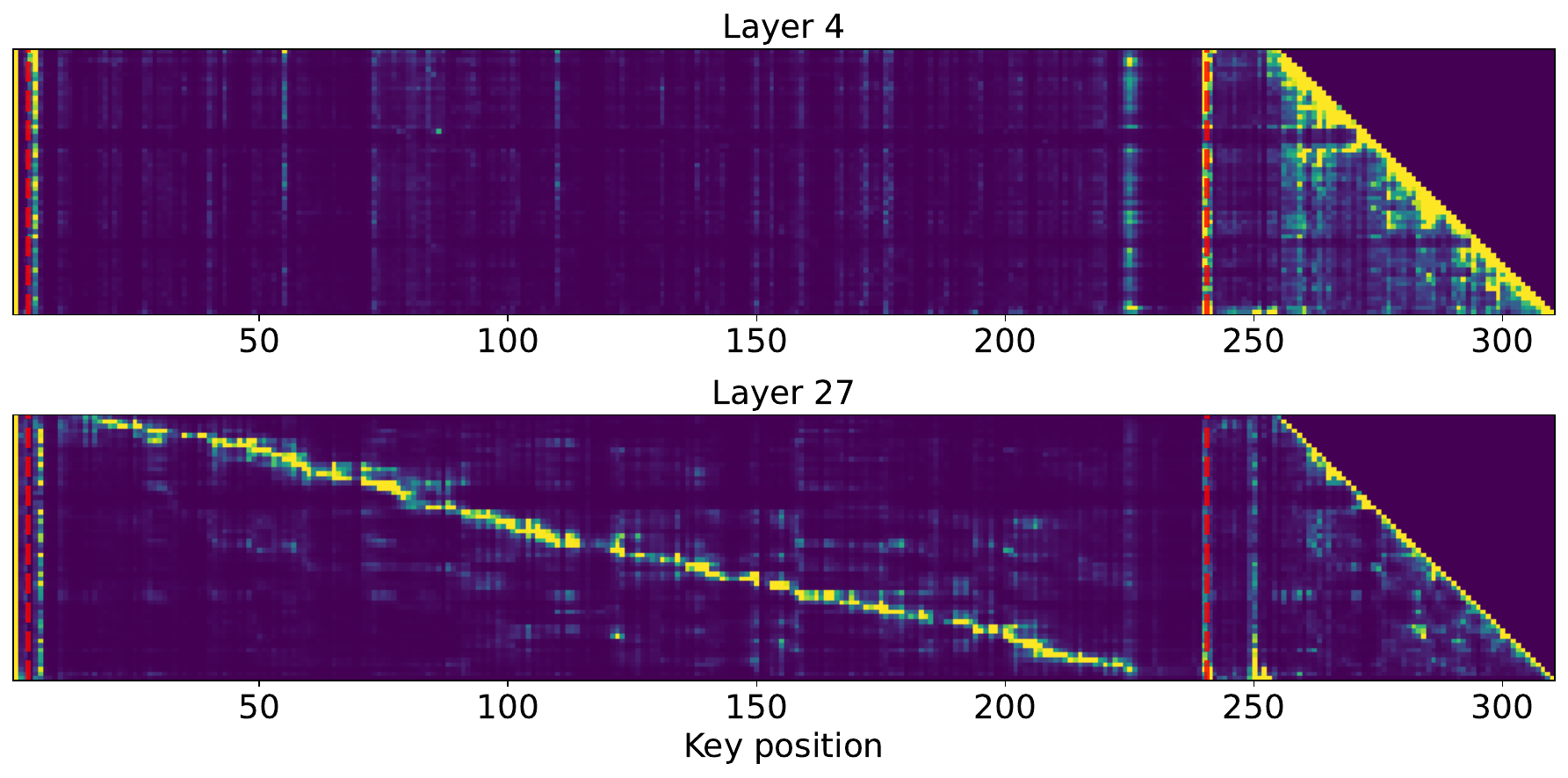}
        \caption{Baseline}
        \label{fig:attn_baseline}
    \end{subfigure}

    \begin{subfigure}{0.98\linewidth}
        \centering
        \includegraphics[width=\linewidth]{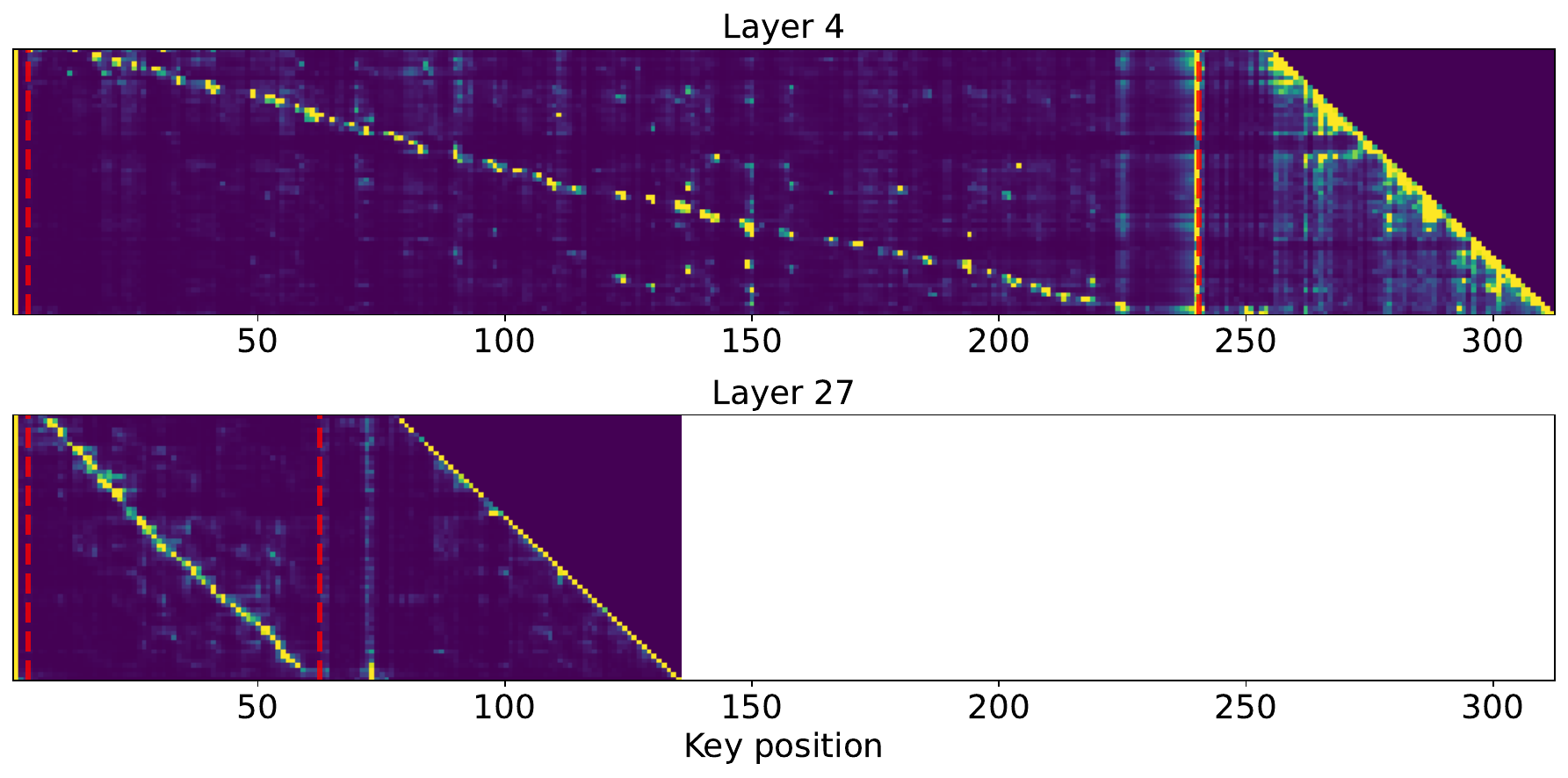}
        \caption{SpeechKV ($R{=}4$)}
        \label{fig:attn_kv}
    \end{subfigure}

\caption{Attention maps of the Baseline and SpeechKV ($R=4$) at an early layer (4) and a deep layer (27) on an ASR example. Each row shows the attention scores from a generated token over the KV cache positions. The region between red dashed lines indicates speech key positions.}
    
    \label{fig:attention}
\end{figure}

As a preliminary analysis, we measure local similarity before any compression training. As shown in Figure~\ref{fig:local_similarity_comparison}, the baseline's local similarity rises steeply through the first four layers, indicating that the model is actively aggregating local acoustic features into contextualized representations. Around layer~5 the curve transitions into a plateau regime, suggesting that local aggregation has largely completed and temporal redundancy has naturally emerged. Based on this observation, we select layer~5 as the compression starting point for our main experiments in Section~\ref{subsec:main_results}.

We further compare the baseline model with SpeechKV after training. As shown in Figure~\ref{fig:local_similarity_comparison}, at $R{=}4$ the local similarity closely follows the baseline's natural rising trend, suggesting that the compression aligns with the aggregation the model would perform on its own, which may explain why $R{=}4$ matches the uncompressed baseline. Although learned pooling is a simple compression operator, the similarity pattern confirms that it does not substantially alter the LLM's internal behavior at this ratio. At $R{=}8$, however, the similarity is elevated across all layers, indicating that the model must restructure its representations to accommodate the more aggressive compression, consistent with the slight quality degradation observed in out-of-domain metrics in Table~\ref{tab:main}.

Table~\ref{tab:layer_ablation} reports the ablation results for different starting layers $l_0$. On OpenASR, layers~5 and~7 yield comparable WER, indicating that the representation is already sufficiently aggregated by layer~5 and remains stable thereafter. However, the more challenging In-house benchmark reveals a clear optimum at layer~5, with both earlier ($l_0{=}3$) and later ($l_0{=}7$) entry points suffering notable degradation in EER and WER. 
This result may reflect the intrinsic layer-wise behavior of Qwen3-1.7B observed in Figure~\ref{fig:local_similarity_comparison}. Starting at layer~3 compresses representations before local similarity has converged, while layer~7 may have already begun to specialize for high-level processing. 
Extending this analysis to further understand the layer-wise dynamics within speech LLMs poses an interesting direction for future work.

\subsection{Attention Analysis}
\label{subsec:attention_analysis}

Figure~\ref{fig:attention} visualizes the attention maps from the generated ASR tokens to the key cache at an early layer~(4) and a deep layer~(27) for both baseline and SpeechKV.
Each row corresponds to a generated token, and brighter values indicate stronger attention weights.
The diagonal band reflects strong speech-text alignment where each text token primarily attends to the speech region it transcribes.

In the baseline (Fig.~\ref{fig:attn_baseline}), attention at the early layer is diffuse, suggesting that the model is still aggregating local acoustic features and has not yet formed explicit speech-text correlations.
By layer~27, a clear monotonic pattern emerges where each text token attends to its corresponding speech span, indicating that a Speech LLM naturally develops speech-text alignment as it progresses through the network.
However, because adjacent key positions are highly similar at deep layers (see Section~\ref{subsec:redundancy}), attention weight is spread across these near-duplicate positions rather than concentrated on the most informative ones.

With SpeechKV (Fig.~\ref{fig:attn_kv}), the behaviour differs at both depths.
At the early layer, a monotonic alignment already appears even though compression has not yet been applied at this layer.
A possible explanation is that, when deep-layer key sequences are shortened during training, the model adapts by consolidating acoustic information earlier in the network, resulting in speech-text correspondence that emerges at shallower layers.
At the deep layer, the compressed key sequences produce sharper, more concentrated attention peaks compared to the baseline, as redundant positions have been merged and each remaining key carries more distinctive information.
This is consistent with the improved recognition accuracy discussed in Section~\ref{subsec:main_results}.

Taken together, these observations suggest that training with SpeechKV reshapes the internal representations across layers.
The early layers take on a greater role in establishing cross-modal correspondence, while the deeper layers operate more efficiently over a compact, de-duplicated key representation that focuses attention on the most informative positions.

\subsection{Decoding Speed}

\begin{figure}[t]
    \centering
    \includegraphics[width=\linewidth]{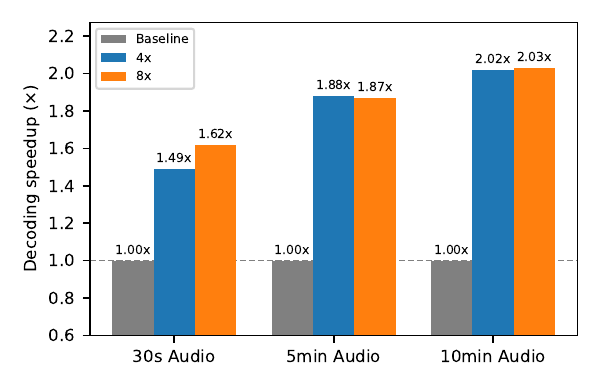}
    \caption{End-to-end decoding speedup at each decoding step on an A100 80\,GB GPU with vLLM. Batch sizes are maximized to saturate GPU memory for each scenario.}
    \label{fig:speed}
\end{figure}

At inference time, SpeechKV directly reduces the KV cache memory by $2.8\times$ and $4\times$ at $R{=}4$ and $R{=}8$, respectively, and the shorter key sequences also reduce the attention computation at each decoding step.

Figure~\ref{fig:speed} reports the end-to-end decoding speedup measured on an A100 80\,GB GPU with vLLM~\cite{kwon2023efficient} across different audio lengths and compression ratios.
The speedup scales with audio length because the speech KV cache occupies a larger fraction of the total attention cost for longer inputs.
At $R{=}4$, the per-step speedup grows from $1.49\times$ for 30-second utterances to $2.02\times$ for 10-minute audio.
Doubling the compression to $R{=}8$, however, yields only a modest additional gain, as the feed-forward layers and text KV cache, which are unaffected by SpeechKV, become the bottleneck once the speech KV cache is sufficiently small.

Considering both speed and recognition quality (Table~\ref{tab:main}), $R{=}4$ offers the most practical operating point that matches the uncompressed baseline in accuracy while delivering $1.49$ to $2\times$ faster decoding that scales with audio length.

\section{Conclusion}
\label{sec:conclusion}

We presented SpeechKV, a method that compresses the speech KV cache at an intermediate LLM layer using learned pooling, deferring the reduction of speech sequences into the LLM itself rather than at the adapter level. By preserving full-resolution queries and the residual stream, SpeechKV allows the LLM's pretrained parameters to govern which acoustic features to retain while substantially reducing the per-step attention cost during decoding.

Experiments on a Qwen3-1.7B-based Speech LLM trained on 71K hours show that compressing speech KV cache to text-level granularity ($R{=}4$) still matches or even slightly improves the uncompressed baseline, with relative gains of up to 6.6\% on entity recognition, while achieving at least $1.49\times$ decoding speedup that scales with audio length. 
Analysis of attention patterns and local similarity reveals that SpeechKV reshapes the layer-wise behavior: early layers establish cross-modal correspondence sooner, while deeper layers attend more sharply over the de-duplicated key sequences, acting as a structural regularization for enhancing speech-text alignment.

Future work includes extending SpeechKV to other speech tasks beyond ASR, investigating adaptive compression ratios across layers, and further analyzing the layer-wise dynamics within Speech LLMs.

\section{Generative AI Use Disclosure}
Generative AI tools were used solely to assist with language polishing and editing of this manuscript. All technical content, experimental design, results, and analysis were produced by the authors.

\newpage
\bibliographystyle{ieeetr}
\bibliography{main}

@INPROCEEDINGS{desta2,
  author={Lu, Ke-Han and others},
  booktitle={{IEEE International Conference on Acoustics, Speech and Signal Processing (ICASSP)}}, 
  title={{Developing Instruction-Following Speech Language Model Without Speech Instruction-Tuning Data}}, 
  year={2025},
  volume={},
  number={},
  pages={1-5},
  doi={10.1109/ICASSP49660.2025.10889444}}

@inproceedings{
tang2024salmonn,
title={{{SALMONN}}: Towards Generic Hearing Abilities for Large Language Models},
author={Changli Tang and Wenyi Yu and Guangzhi Sun and Xianzhao Chen and Tian Tan and Wei Li and Lu Lu and Zejun MA and Chao Zhang},
booktitle={{The Twelfth International Conference on Learning Representations}},
year={2024},
}

@inproceedings{lu24c_interspeech,
  title     = {{DeSTA}: Enhancing Speech Language Models through Descriptive Speech-Text Alignment},
  author    = {Ke-Han Lu and Zhehuai Chen and Szu-Wei Fu and He Huang and Boris Ginsburg and Yu-Chiang Frank Wang and Hung-Yi Lee},
  year      = {2024},
  booktitle = {Interspeech 2024},
  pages     = {4159--4163},
  doi       = {10.21437/Interspeech.2024-457},
  issn      = {2958-1796},
}

@ARTICLE{lu2025desta25,
author={Lu, Ke-Han and Chen, Zhehuai and Fu, Szu-Wei and Yang, Chao-Han Huck and Huang, Sung-Feng and Yang, Chih-Kai and Yu, Chee-En and Chen, Chun-Wei and Chen, Wei-Chih and Huang, Chien-yu and Lin, Yi-Cheng and Lin, Yu-Xiang and Fu, Chi-An and Kuan, Chun-Yi and Ren, Wenze and Chen, Xuanjun and Huang, Wei-Ping and Hu, En-Pei and Lin, Tzu-Quan and Wu, Yuan-Kuei and Huang, Kuan-Po and Huang, Hsiao-Ying and Chou, Huang-Cheng and Chang, Kai-Wei and Chiang, Cheng-Han and Ginsburg, Boris and Wang, Yu-Chiang Frank and Lee, Hung-yi},
journal={IEEE Transactions on Audio, Speech and Language Processing}, 
title={DeSTA2.5-Audio: Toward General-Purpose Large Audio Language Model with Self-Generated Cross-Modal Alignment}, 
year={2026},
volume={},
number={},
pages={1-16},
doi={10.1109/TASLPRO.2026.3675792}}

@article{hu2024wavllm,
  title={{Wavllm: Towards robust and adaptive speech large language model}},
  author={Hu, Shujie and others},
  journal={arXiv preprint arXiv:2404.00656},
  year={2024}
}

@article{held2024distilling,
  title={{Distilling an end-to-end voice assistant without instruction training data}},
  author={Held, William and Li, Ella and Ryan, Michael and Shi, Weiyan and Zhang, Yanzhe and Yang, Diyi},
  journal={arXiv preprint arXiv:2410.02678},
  year={2024}
}

@article{peng2023instruction,
  title={{Instruction tuning with gpt-4}},
  author={Peng, Baolin and Li, Chunyuan and He, Pengcheng and Galley, Michel and Gao, Jianfeng},
  journal={arXiv preprint arXiv:2304.03277},
  year={2023}
}

@article{yang2025qwen3,
  title={Qwen3 technical report},
  author={Yang, An and Li, Anfeng and Yang, Baosong and Zhang, Beichen and Hui, Binyuan and Zheng, Bo and Yu, Bowen and Gao, Chang and Huang, Chengen and Lv, Chenxu and others},
  journal={arXiv preprint arXiv:2505.09388},
  year={2025}
}

@inproceedings{kwon2023efficient,
  title={Efficient Memory Management for Large Language Model Serving with PagedAttention},
  author={Woosuk Kwon and Zhuohan Li and Siyuan Zhuang and Ying Sheng and Lianmin Zheng and Cody Hao Yu and Joseph E. Gonzalez and Hao Zhang and Ion Stoica},
  booktitle={Proceedings of the ACM SIGOPS 29th Symposium on Operating Systems Principles},
  year={2023}
}

@article{dubey2024llama,
  title={{The llama 3 herd of models}},
  author={Dubey, Abhimanyu and Jauhri, Abhinav and Pandey, Abhinav and Kadian, Abhishek and Al-Dahle, Ahmad and Letman, Aiesha and Mathur, Akhil and Schelten, Alan and Yang, Amy and Fan, Angela and others},
  journal={arXiv preprint arXiv:2407.21783},
  year={2024}
}

@article{yang2024qwen2technicalreport,
      title={{Qwen2 Technical Report}}, 
      author={Yang, An and others},
      journal={arXiv preprint arXiv:2412.15115},
    year={2024}
}

@inproceedings{radford2023robust,
  title={{Robust speech recognition via large-scale weak supervision}},
  author={Radford, Alec and Kim, Jong Wook and Xu, Tao and Brockman, Greg and McLeavey, Christine and Sutskever, Ilya},
  booktitle={{International conference on machine learning}},
  year={2023},
  organization={PMLR}
}

@article{abdin2024phi,
  title={{Phi-4 technical report}},
  author={Abdin, Marah and Aneja, Jyoti and Behl, Harkirat and Bubeck, S{\'e}bastien and Eldan, Ronen and Gunasekar, Suriya and Harrison, Michael and Hewett, Russell J and Javaheripi, Mojan and Kauffmann, Piero and others},
  journal={arXiv preprint arXiv:2412.08905},
  year={2024}
}

@INPROCEEDINGS{10389705,
  author={Wu, Jian and Gaur, Yashesh and Chen, Zhuo and Zhou, Long and Zhu, Yimeng and Wang, Tianrui and Li, Jinyu and Liu, Shujie and Ren, Bo and Liu, Linquan and Wu, Yu},
  booktitle={2023 IEEE Automatic Speech Recognition and Understanding Workshop (ASRU)}, 
  title={On Decoder-Only Architecture For Speech-to-Text and Large Language Model Integration}, 
  year={2023},
  volume={},
  number={},
  pages={1-8},
  doi={10.1109/ASRU57964.2023.10389705}}

@article{abouelenin2025phi,
  title={Phi-4-mini technical report: Compact yet powerful multimodal language models via mixture-of-loras},
  author={Abouelenin, Abdelrahman and Ashfaq, Atabak and Atkinson, Adam and Awadalla, Hany and Bach, Nguyen and Bao, Jianmin and Benhaim, Alon and Cai, Martin and Chaudhary, Vishrav and Chen, Congcong and others},
  journal={arXiv preprint arXiv:2503.01743},
  year={2025}
}

@inproceedings{commonvoice:2020,   author = {Ardila, R. and Branson, M. and Davis, K. and Henretty, M. and Kohler, M. and Meyer, J. and Morais, R. and Saunders, L. and Tyers, F. M. and Weber, G.},   title = {Common Voice: A Massively-Multilingual Speech Corpus},   booktitle = {Proceedings of the 12th Conference on Language Resources and Evaluation (LREC 2020)},   pages = {4211--4215},   year = 2020 }

@inproceedings{panayotov2015librispeech,
  title={Librispeech: an asr corpus based on public domain audio books},
  author={Panayotov, Vassil and Chen, Guoguo and Povey, Daniel and Khudanpur, Sanjeev},
  booktitle={2015 IEEE international conference on acoustics, speech and signal processing (ICASSP)},
  pages={5206--5210},
  year={2015},
  organization={IEEE}
}

@InProceedings{pmlr-v202-li23q,
  title = 	 {{BLIP}-2: Bootstrapping Language-Image Pre-training with Frozen Image Encoders and Large Language Models},
  author =       {Li, Junnan and Li, Dongxu and Savarese, Silvio and Hoi, Steven},
  booktitle = 	 {Proceedings of the 40th International Conference on Machine Learning},
  pages = 	 {19730--19742},
  year = 	 {2023},
  volume = 	 {202},
  publisher =    {PMLR},
  url = 	 {}
}

@inproceedings{hulora,
  title={{LoRA: Low-Rank Adaptation of Large Language Models}},
  author={Hu, Edward J and Wallis, Phillip and Allen-Zhu, Zeyuan and Li, Yuanzhi and Wang, Shean and Wang, Lu and Chen, Weizhu and others},
  booktitle={{International Conference on Learning Representations}}
}

@article{arora2025landscapespokenlanguagemodels,
  title={On the landscape of spoken language models: A comprehensive survey},
  author={Arora, Siddhant and Chang, Kai-Wei and Chien, Chung-Ming and Peng, Yifan and Wu, Haibin and Adi, Yossi and Dupoux, Emmanuel and Lee, Hung-Yi and Livescu, Karen and Watanabe, Shinji},
  journal={arXiv preprint arXiv:2504.08528},
  year={2025}
}

@ARTICLE{11077996,
  author={Fan, Ruchao and Ren, Bo and Hu, Yuxuan and Zhao, Rui and Liu, Shujie and Li, Jinyu},
  journal={IEEE Journal of Selected Topics in Signal Processing}, 
  title={AlignFormer: Modality Matching Can Achieve Better Zero-Shot Instruction-Following Speech-LLM}, 
  year={2025},
  volume={19},
  number={7},
  pages={1329-1337},
  doi={10.1109/JSTSP.2025.3588378}}

@INPROCEEDINGS{10447605,
  author={Fathullah, Yassir and Wu, Chunyang and Lakomkin, Egor and Jia, Junteng and Shangguan, Yuan and Li, Ke and Guo, Jinxi and Xiong, Wenhan and Mahadeokar, Jay and Kalinli, Ozlem and Fuegen, Christian and Seltzer, Mike},
  booktitle={ICASSP 2024 - 2024 IEEE International Conference on Acoustics, Speech and Signal Processing (ICASSP)}, 
  title={Prompting Large Language Models with Speech Recognition Abilities}, 
  year={2024},
  volume={},
  number={},
  pages={13351-13355},
  doi={10.1109/ICASSP48485.2024.10447605}}

@INPROCEEDINGS{10445874,
  author={Yu, Wenyi and Tang, Changli and Sun, Guangzhi and Chen, Xianzhao and Tan, Tian and Li, Wei and Lu, Lu and Ma, Zejun and Zhang, Chao},
  booktitle={ICASSP 2024 - 2024 IEEE International Conference on Acoustics, Speech and Signal Processing (ICASSP)}, 
  title={Connecting Speech Encoder and Large Language Model for ASR}, 
  year={2024},
  volume={},
  number={},
  pages={12637-12641},
  doi={10.1109/ICASSP48485.2024.10445874}}

@inproceedings{
nawrot2024dynamic,
title={Dynamic Memory Compression: Retrofitting {LLM}s for Accelerated Inference},
author={Piotr Nawrot and Adrian {\L}a{\'n}cucki and Marcin Chochowski and David Tarjan and Edoardo Ponti},
booktitle={Forty-first International Conference on Machine Learning},
year={2024},
url={https://openreview.net/forum?id=tDRYrAkOB7}
}

@article{wang2024model,
  title={Model tells you where to merge: Adaptive kv cache merging for llms on long-context tasks},
  author={Wang, Zheng and Jin, Boxiao and Yu, Zhongzhi and Zhang, Minjia},
  journal={arXiv preprint arXiv:2407.08454},
  year={2024}
}

@article{
li2025a,
title={A Survey on Large Language Model Acceleration based on {KV} Cache Management},
author={Haoyang LI and Yiming Li and Anxin Tian and Tianhao Tang and Zhanchao Xu and Xuejia Chen and Nicole HU and Wei Dong and Li Qing and Lei Chen},
journal={Transactions on Machine Learning Research},
issn={2835-8856},
year={2025},
url={https://openreview.net/forum?id=z3JZzu9EA3},
note={}
}

@inproceedings{NEURIPS2023_6ceefa7b,
 author = {Zhang, Zhenyu and Sheng, Ying and Zhou, Tianyi and Chen, Tianlong and Zheng, Lianmin and Cai, Ruisi and Song, Zhao and Tian, Yuandong and R\'{e}, Christopher and Barrett, Clark and Wang, Zhangyang "Atlas" and Chen, Beidi},
 booktitle = {Advances in Neural Information Processing Systems},
 editor = {A. Oh and T. Naumann and A. Globerson and K. Saenko and M. Hardt and S. Levine},
 pages = {34661--34710},
 publisher = {Curran Associates, Inc.},
 title = {H2O: Heavy-Hitter Oracle for Efficient Generative Inference of Large Language Models},
 url = {https://proceedings.neurips.cc/paper_files/paper/2023/file/6ceefa7b15572587b78ecfcebb2827f8-Paper-Conference.pdf},
 volume = {36},
 year = {2023}
}

@inproceedings{NEURIPS2024_28ab4182,
 author = {Li, Yuhong and Huang, Yingbing and Yang, Bowen and Venkitesh, Bharat and Locatelli, Acyr and Ye, Hanchen and Cai, Tianle and Lewis, Patrick and Chen, Deming},
 booktitle = {Advances in Neural Information Processing Systems},
 doi = {10.52202/079017-0722},
 editor = {A. Globerson and L. Mackey and D. Belgrave and A. Fan and U. Paquet and J. Tomczak and C. Zhang},
 pages = {22947--22970},
 publisher = {Curran Associates, Inc.},
 title = {SnapKV: LLM Knows What You are Looking for Before Generation},
 url = {https://proceedings.neurips.cc/paper_files/paper/2024/file/28ab418242603e0f7323e54185d19bde-Paper-Conference.pdf},
 volume = {37},
 year = {2024}
}

@article{sun2026speech,
  title={Speech-XL: Towards Long-Form Speech Understanding in Large Speech Language Models},
  author={Sun, Haoqin and Lyu, Chenyang and Zhao, Shiwan and Ni, Xuanfan and Kong, Xiangyu and Wang, Longyue and Luo, Weihua and Qin, Yong},
  journal={arXiv preprint arXiv:2602.05373},
  year={2026}
}

@article{deng2026speech,
  title={Speech LLMs are Contextual Reasoning Transcribers},
  author={Deng, Keqi and Fan, Ruchao and Ren, Bo and Wang, Yiming and Li, Jinyu},
  journal={arXiv preprint arXiv:2604.00610},
  year={2026}
}

@INPROCEEDINGS{10888940,
  author={Zhou, Wei and Jia, Junteng and Sari, Leda and Mahadeokar, Jay and Kalinli, Ozlem},
  booktitle={ICASSP 2025 - 2025 IEEE International Conference on Acoustics, Speech and Signal Processing (ICASSP)}, 
  title={CJST: CTC Compressor based Joint Speech and Text Training for Decoder-Only ASR}, 
  year={2025},
  volume={},
  number={},
  pages={1-5},
  doi={10.1109/ICASSP49660.2025.10888940}}

@inproceedings{deng-etal-2025-wav2prompt,
    title = "{W}av2{P}rompt: End-to-End Speech Prompt Learning and Task-based Fine-tuning for Text-based {LLM}s",
    author = "Deng, Keqi  and
      Sun, Guangzhi  and
      Woodland, Phil",
    editor = "Chiruzzo, Luis  and
      Ritter, Alan  and
      Wang, Lu",
    booktitle = "Proceedings of the 2025 Conference of the Nations of the Americas Chapter of the Association for Computational Linguistics: Human Language Technologies (Volume 1: Long Papers)",
    month = apr,
    year = "2025",
    address = "Albuquerque, New Mexico",
    publisher = "Association for Computational Linguistics",
    url = "https://aclanthology.org/2025.naacl-long.354/",
    doi = "10.18653/v1/2025.naacl-long.354",
    pages = "6940--6956",
    ISBN = "979-8-89176-189-6"
}

@article{verdini2024connect,
  title={How to connect speech foundation models and large language models? what matters and what does not},
  author={Verdini, Francesco and Melucci, Pierfrancesco and Perna, Stefano and Cariaggi, Francesco and Gaido, Marco and Papi, Sara and Mazurek, Szymon and Kasztelnik, Marek and Bentivogli, Luisa and Brati{\`e}res, S{\'e}bastien and others},
  journal={arXiv preprint arXiv:2409.17044},
  year={2024}
}

@article{ma2024embarrassingly,
  title={An Embarrassingly Simple Approach for LLM with Strong ASR Capacity},
  author={Ma, Ziyang and Yang, Guanrou and Yang, Yifan and Gao, Zhifu and Wang, Jiaming and Du, Zhihao and Yu, Fan and Chen, Qian and Zheng, Siqi and Zhang, Shiliang and others},
  journal={arXiv preprint arXiv:2402.08846},
  year={2024}
}

@inproceedings{chevalier-etal-2023-adapting,
    title = "Adapting Language Models to Compress Contexts",
    author = "Chevalier, Alexis  and
      Wettig, Alexander  and
      Ajith, Anirudh  and
      Chen, Danqi",
    editor = "Bouamor, Houda  and
      Pino, Juan  and
      Bali, Kalika",
    booktitle = "Proceedings of the 2023 Conference on Empirical Methods in Natural Language Processing",
    month = dec,
    year = "2023",
    address = "Singapore",
    publisher = "Association for Computational Linguistics",
    url = "https://aclanthology.org/2023.emnlp-main.232/",
    doi = "10.18653/v1/2023.emnlp-main.232",
    pages = "3829--3846"
}

@inproceedings{ICLR2024_5e5fd18f,
 author = {Xiao, Guangxuan and Tian, Yuandong and Chen, Beidi and Han, Song and Lewis, Mike },
 booktitle = {International Conference on Learning Representations},
 editor = {B. Kim and Y. Yue and S. Chaudhuri and K. Fragkiadaki and M. Khan and Y. Sun},
 pages = {21875--21895},
 title = {Efficient Streaming Language Models with Attention Sinks},
 url = {https://proceedings.iclr.cc/paper_files/paper/2024/file/5e5fd18f863cbe6d8ae392a93fd271c9-Paper-Conference.pdf},
 volume = {2024},
 year = {2024}
}

@inproceedings{NEURIPS2024_028fcbcf,
 author = {Hooper, Coleman and Kim, Sehoon and Mohammadzadeh, Hiva and Mahoney, Michael W. and Shao, Yakun Sophia and Keutzer, Kurt and Gholami, Amir},
 booktitle = {Advances in Neural Information Processing Systems},
 doi = {10.52202/079017-0040},
 editor = {A. Globerson and L. Mackey and D. Belgrave and A. Fan and U. Paquet and J. Tomczak and C. Zhang},
 pages = {1270--1303},
 publisher = {Curran Associates, Inc.},
 title = {KVQuant: Towards 10 Million Context Length LLM Inference with KV Cache Quantization},
 url = {https://proceedings.neurips.cc/paper_files/paper/2024/file/028fcbcf85435d39a40c4d61b42c99a4-Paper-Conference.pdf},
 volume = {37},
 year = {2024}
}

@article{mohapatra2026speechmapper,
  title={SpeechMapper: Speech-to-text Embedding Projector for LLMs},
  author={Mohapatra, Biswesh and Boito, Marcely Zanon and Calapodescu, Ioan},
  journal={arXiv preprint arXiv:2601.20417},
  year={2026}
}

@article{Pratap2020MLSAL,
  title={MLS: A Large-Scale Multilingual Dataset for Speech Research},
  author={Vineel Pratap and Qiantong Xu and Anuroop Sriram and Gabriel Synnaeve and Ronan Collobert},
  journal={ArXiv},
  year={2020},
  volume={abs/2012.03411}
}

@inproceedings{GigaSpeech2021,
  title     = {{GigaSpeech: An Evolving, Multi-Domain ASR Corpus with 10,000 Hours of Transcribed Audio}},
  author    = {Guoguo Chen and Shuzhou Chai and Guan-Bo Wang and Jiayu Du and Wei-Qiang Zhang and Chao Weng and Dan Su and Daniel Povey and Jan Trmal and Junbo Zhang and Mingjie Jin and Sanjeev Khudanpur and Shinji Watanabe and Shuaijiang Zhao and Wei Zou and Xiangang Li and Xuchen Yao and Yongqing Wang and Zhao You and Zhiyong Yan},
  year      = {2021},
  booktitle = {{Interspeech 2021}},
  pages     = {3670--3674},
  doi       = {10.21437/Interspeech.2021-1965},
  issn      = {2958-1796},
}

@ARTICLE{2021arXiv210402014O,
       author = {{O'Neill}, Patrick K. and {Lavrukhin}, Vitaly and {Majumdar}, 
       Somshubra and {Noroozi}, Vahid and {Zhang}, Yuekai and {Kuchaiev}, Oleksii and {Balam}, 
       Jagadeesh and {Dovzhenko}, Yuliya and {Freyberg}, Keenan and {Shulman}, Michael D. and {Ginsburg}, 
       Boris and {Watanabe}, Shinji and {Kucsko}, Georg},
        title = "{SPGISpeech: 5,000 hours of transcribed financial audio for fully formatted end-to-end speech recognition}",
      journal = {arXiv e-prints},
         year = 2021,
        month = apr,
          eid = {arXiv:2104.02014},
        pages = {arXiv:2104.02014},
archivePrefix = {arXiv},
       eprint = {2104.02014},
 primaryClass = {cs.CL},
       adsurl = {https://ui.adsabs.harvard.edu/abs/2021arXiv210402014O}
}

@inproceedings{wang-etal-2021-voxpopuli,
    title = "{V}ox{P}opuli: A Large-Scale Multilingual Speech Corpus for Representation Learning, Semi-Supervised Learning and Interpretation",
    author = "Wang, Changhan  and
      Riviere, Morgane  and
      Lee, Ann  and
      Wu, Anne  and
      Talnikar, Chaitanya  and
      Haziza, Daniel  and
      Williamson, Mary  and
      Pino, Juan  and
      Dupoux, Emmanuel",
    editor = "Zong, Chengqing  and
      Xia, Fei  and
      Li, Wenjie  and
      Navigli, Roberto",
    booktitle = "Proceedings of the 59th Annual Meeting of the Association for Computational Linguistics and the 11th International Joint Conference on Natural Language Processing (Volume 1: Long Papers)",
    month = aug,
    year = "2021",
    address = "Online",
    publisher = "Association for Computational Linguistics",
    url = "https://aclanthology.org/2021.acl-long.80/",
    doi = "10.18653/v1/2021.acl-long.80",
    pages = "993--1003"
}

@inproceedings{hernandez2018ted,
  title={TED-LIUM 3: Twice as much data and corpus repartition for experiments on speaker adaptation},
  author={Hernandez, Fran{\c{c}}ois and Nguyen, Vincent and Ghannay, Sahar and Tomashenko, Natalia and Esteve, Yannick},
  booktitle={Speech and Computer: 20th International Conference, SPECOM 2018, Leipzig, Germany, September 18--22, 2018, Proceedings 20},
  pages={198--208},
  year={2018},
  organization={Springer}
}

@misc{delrio2022earnings22,
      title={"Earnings-22: A Practical Benchmark for Accents in the Wild"}, 
      author={Miguel Del Rio and Peter Ha and Quinten McNamara and Corey Miller and Shipra Chandra},
      year={2022},
      eprint={2203.15591},
      archivePrefix={arXiv},
      primaryClass={cs.CL}
}

@article{fleurs2022arxiv,
  title = {FLEURS: Few-shot Learning Evaluation of Universal Representations of Speech},
  author = {Conneau, Alexis and Ma, Min and Khanuja, Simran and Zhang, Yu and Axelrod, Vera and Dalmia, Siddharth and Riesa, Jason and Rivera, Clara and Bapna, Ankur},
  journal={arXiv preprint arXiv:2205.12446},
  url = {https://arxiv.org/abs/2205.12446},
  year = {2022},
}

@InProceedings{10.1007/11677482_3,
author="Carletta, Jean
and Ashby, Simone
and Bourban, Sebastien
and Flynn, Mike
and Guillemot, Mael
and Hain, Thomas
and Kadlec, Jaroslav
and Karaiskos, Vasilis
and Kraaij, Wessel
and Kronenthal, Melissa
and Lathoud, Guillaume
and Lincoln, Mike
and Lisowska, Agnes
and McCowan, Iain
and Post, Wilfried
and Reidsma, Dennis
and Wellner, Pierre",
editor="Renals, Steve
and Bengio, Samy",
title="The AMI Meeting Corpus: A Pre-announcement",
booktitle="Machine Learning for Multimodal Interaction",
year="2006",
publisher="Springer Berlin Heidelberg",
address="Berlin, Heidelberg",
pages="28--39",
isbn="978-3-540-32550-5"
}

@misc{kang2023libriheavy,
      title={Libriheavy: a 50,000 hours ASR corpus with punctuation casing and context}, 
      author={Wei Kang and Xiaoyu Yang and Zengwei Yao and Fangjun Kuang and Yifan Yang and Liyong Guo and Long Lin and Daniel Povey},
      year={2023},
      eprint={2309.08105},
      archivePrefix={arXiv},
      primaryClass={eess.AS}
}

@misc{srivastav2025openasrleaderboardreproducible,
      title={Open ASR Leaderboard: Towards Reproducible and Transparent Multilingual and Long-Form Speech Recognition Evaluation}, 
      author={Vaibhav Srivastav and Steven Zheng and Eric Bezzam and Eustache Le Bihan and Nithin Koluguri and Piotr Żelasko and Somshubra Majumdar and Adel Moumen and Sanchit Gandhi},
      year={2026},
      eprint={2510.06961},
      archivePrefix={arXiv},
      primaryClass={cs.CL},
      url={https://arxiv.org/abs/2510.06961}, 
}

@inproceedings{gulati20_interspeech,
  title     = {{Conformer: Convolution-augmented Transformer for Speech Recognition}},
  author    = {Anmol Gulati and James Qin and Chung-Cheng Chiu and Niki Parmar and Yu Zhang and Jiahui Yu and Wei Han and Shibo Wang and Zhengdong Zhang and Yonghui Wu and Ruoming Pang},
  year      = {2020},
  booktitle = {{Interspeech 2020}},
  pages     = {5036--5040},
  doi       = {10.21437/Interspeech.2020-3015},
  issn      = {2958-1796},
}

@inproceedings{10.1145/1143844.1143891,
author = {Graves, Alex and Fern\'{a}ndez, Santiago and Gomez, Faustino and Schmidhuber, J\"{u}rgen},
title = {Connectionist temporal classification: labelling unsegmented sequence data with recurrent neural networks},
year = {2006},
isbn = {1595933832},
publisher = {Association for Computing Machinery},
address = {New York, NY, USA},
url = {https://doi.org/10.1145/1143844.1143891},
doi = {10.1145/1143844.1143891},
booktitle = {Proceedings of the 23rd International Conference on Machine Learning},
pages = {369–376},
numpages = {8},
location = {Pittsburgh, Pennsylvania, USA},
series = {ICML '06}
}

@inproceedings{dong2020cif,
  title={Cif: Continuous integrate-and-fire for end-to-end speech recognition},
  author={Dong, Linhao and Xu, Bo},
  booktitle={ICASSP 2020-2020 IEEE International Conference on Acoustics, Speech and Signal Processing (ICASSP)},
  pages={6079--6083},
  year={2020},
  organization={IEEE}
}

@inproceedings{3737916.3742359,
author = {Liu, Akide and Liu, Jing and Pan, Zizheng and He, Yefei and Haffari, Gholamreza and Zhuang, Bohan},
title = {MiniCache: KV cache compression in depth dimension for large language models},
year = {2024},
isbn = {9798331314385},
publisher = {Curran Associates Inc.},
address = {Red Hook, NY, USA},
booktitle = {Proceedings of the 38th International Conference on Neural Information Processing Systems},
articleno = {4443},
numpages = {35},
location = {Vancouver, BC, Canada},
series = {NIPS '24}
}

\end{document}